\PassOptionsToPackage{usenames,dvipsnames}{xcolor}

\documentclass[sigconf, research, screen, nonacm]{acmart}
\usepackage{enumitem}
\usepackage{multirow}
\usepackage{mathpazo} 

\newcommand{\FrameworkName}{$\mathcal{A}^2\mathcal{C}$}

\newcommand{\WarningMark}{\raisebox{.20ex}{{\fontencoding{U}\fontfamily{futs}\selectfont\char 49 \relax}}}

\usepackage{soul}

\usepackage{tcolorbox}
\tcbuselibrary{listings,theorems}
\usepackage{lineno}
\newcommand{\sh}[1]{\textcolor{RoyalBlue}{Sha: #1}}

\newcommand\mfont{\fontsize{7.6pt}{18pt}\selectfont}
\newcommand{\CIFAR}{C{\mfont IFAR}}
\newcommand{\MNIST}{M{\mfont NIST}}
\newcommand{\fMNIST}{FM{\mfont NIST}}
\newcommand{\KDDCup}{K{\mfont DD}Cup{\mfont 99}}
\hypersetup{
    colorlinks=true,
    urlcolor=blue,
    citecolor=Magenta,
    linkcolor=red,
}

\usepackage{array}
\usepackage{multirow}
\usepackage{ragged2e}
\usepackage{booktabs}

\AtBeginDocument{%
  \providecommand\BibTeX{{%
    \normalfont B\kern-0.5em{\scshape i\kern-0.25em b}\kern-0.8em\TeX}}}

\setcopyright{acmcopyright}
\copyrightyear{2018}
\acmYear{2018}
\acmDOI{XXXXXXX.XXXXXXX}

\acmConference[Conference acronym 'XX]{Make sure to enter the correct
  conference title from your rights confirmation emai}{June 03--05,
  2018}{Woodstock, NY}
\acmPrice{15.00}
\acmISBN{978-1-4503-XXXX-X/18/06}

\begin{document}

\title{\FrameworkName: A Modular Multi-stage Collaborative Decision Framework for Human-AI Teams}
\settopmatter{printfolios=true}

\author{Shahroz Tariq, Mohan Baruwal Chhetri, Surya Nepal, Cecile Paris}
\email{{shahroz.tariq, mohan.baruwalchhetri, surya.nepal, cecile.paris}@data61.csiro.au}
\affiliation{%
  \institution{CSIRO's Data61, Australia}
    \streetaddress{}
  \city{}
  \state{}
  \country{}
  \postcode{}
}
\begin{abstract}
  This paper introduces \FrameworkName, a multi-stage collaborative decision framework designed to enable robust decision-making within human-AI teams. Drawing inspiration from concepts such as rejection learning and learning to defer, \FrameworkName~incorporates AI systems trained to recognise uncertainty in their decisions and defer to human experts when needed. Moreover, \FrameworkName~caters to scenarios where even human experts encounter limitations, such as in incident detection and response in cyber Security Operations Centres (SOC). In such scenarios, \FrameworkName~facilitates collaborative explorations, enabling collective resolution of complex challenges. With support for three distinct decision-making modes in human-AI teams: \textit{Automated}, \textit{Augmented}, and \textit{Collaborative}, \FrameworkName~offers a flexible platform for developing effective strategies for human-AI collaboration. By harnessing the strengths of both humans and AI, it significantly improves the efficiency and effectiveness of complex decision-making in dynamic and evolving environments. To validate \FrameworkName's capabilities, we conducted extensive simulative experiments using benchmark datasets. The results clearly demonstrate that all three modes of decision-making can be effectively supported by \FrameworkName. Most notably, collaborative exploration by (simulated) human experts and AI achieves superior performance compared to AI in isolation, underscoring the framework's potential to enhance decision-making within human-AI teams. 
\end{abstract}

\maketitle

\section{Introduction} \label{sec:intro}

The exponential rise in cybercrime, and the substantial financial losses resulting from it, has presented significant problems for organisations worldwide. With cybercrime damages on the rise, reaching a staggering \$8.44 trillion USD in 2022 and projected to soar to \$23.8 trillion USD by 2027~\cite{Cybercrime_damage}, organisations face the urgent need to protect their digital assets against an ever-growing threat. To address this, organisations are increasingly relying on Security Operations Centres (SOCs)~\cite{majid2019success}, which are specialised units responsible for real-time detection, analysis, and response to cybersecurity incidents~\cite{SOC_Survey_PPTGC}. However, the increasing complexity of the digital landscape, coupled with the evolving threat landscape,  the overwhelming influx of security alerts, and the shortage of skilled staff, poses significant challenges for the SOC~\cite{SANS_SOC_Survey}. To counter this, SOCs are increasingly harnessing artificial intelligence (AI) and machine learning (ML) technologies to automate certain tasks and augment others, aiming to reduce the cognitive workload of human analysts. 

In recent times, there has been a growing call, including in the cybersecurity domain~\cite{lyn2019opportunities,gomez2019considerations}, for human-AI collaboration~\cite{paris2021} or hybrid intelligence~\cite{akata2020research} wherein humans and AI collaborate with each other leveraging their complementary strengths and compensating for their weaknesses. This collaborative approach aims to develop more robust and effective solutions to achieve goals that are beyond the reach of either humans or AI individually.

While AI and ML technologies offer numerous advantages to domains such as cybersecurity, including real-time data processing, speed, scalability, pattern recognition, predictive analytics, automation and efficiency, and can help address the critical issue of skills shortage, they also have inherent limitations. ML-based AI systems demonstrate high performance within the boundaries of their training data, but exhibit \textit{brittleness} when encountering novel scenarios or unfamiliar patterns~\cite{woods2016risks}. More importantly, unlike humans who know what they do not know, AI systems lack \textit{self-awareness} of their limitations~\cite{national2021human}. Errors stemming from this inherent lack of self-awareness in AI systems can lead to severe consequences in high-risk environments. In the context of cybersecurity, this can include compromised data integrity, unauthorised access to sensitive information, and the potential for catastrophic security breaches. 

To address this challenge, we draw inspiration from concepts such as \textit{learning to reject}~\cite{chow1970optimum}, \textit{learning to defer}~\cite{madras2018predict} and \textit{human-AI ensembles}~\cite{choudhary2023human}. Learning to reject, also known as rejection learning, represents a form of self-awareness in ML models~\cite{RejectionLearning_Survey_new}, where models are trained to abstain from making decisions when faced with uncertainty. Learning to defer extends this concept by considering decision-making in the presence of an external decision-maker~\cite{madras2018predict}. In this approach, the ML model defers difficult cases to an external decision-maker, often human experts, who can leverage their expertise, as well as additional insights, to arrive at a decision. Human-AI ensembles involve the practice of combining the outcomes of human and AI-based algorithmic decision-makers. Most existing approaches are based on the assumption that at least one of the decision-makers - either Human or AI -- can reach a correct decision. However, in certain circumstances, even highly experienced domain experts may encounter novel and open-ended situations that pose unique challenges without a clear and definitive solution. This is particularly the case in the cybersecurity domain, where new threats emerge and existing ones adapt and evolve. In such scenarios, existing approaches prove insufficient, underscoring the need for more effective \textit{human-AI collaboration} that leverages the strengths of both humans and AI to jointly address complex, open-ended problems and explore potential solutions. 
Motivated by these considerations, this work introduces \FrameworkName~-- a modular multi-stage collaborative decision framework for human-AI teams. The framework supports three distinct decision-making modes in human-AI teams: \textit{automated}, \textit{augmented}, and \textit{collaborative}. In the automated mode, the decision-making is primarily handled by AI systems with no human intervention. In the augmented mode, the AI system defers to the human decision-makers, but may assist by providing relevant information, insights, and recommendations. In the collaborative mode, also referred to as \textit{collaborative exploration}, human experts work alongside AI systems in jointly exploring complex scenarios with no clear-cut solutions. Harnessing the combined capabilities of humans and AI can potentially help with decision-making in such scenarios. 
%
%

Strategically integrating these three modes of decision-making creates a flexible environment in which the level of human-AI collaboration can be tailored to suit specific operational needs. This customisation enables efficient and effective collaboration between human experts and AI systems, harnessing the strengths of both to enhance decision-making capabilities, particularly in the cybersecurity domain, which is marked by its ever-evolving and highly dynamic nature. 



We assess the effectiveness of the \FrameworkName\ framework by evaluating it on benchmark datasets tailored to our specific problem using three different scenarios illustrated in \autoref{fig:Pipeline}. Our carefully designed controlled simulations showcase how \FrameworkName\ can enhance performance by combining human and artificial intelligence. For instance, on the \CIFAR-10 and \KDDCup~dataset, (i) automation achieved 37.38\% and 33.43\%, respectively; (ii) selective deferral to a simulated human expert achieved 64.75\% and 35.18\%, respectively; and (iii) collaborative exploration achieved performance gains of up to 81.24\% and 96.24\%, respectively, demonstrating that \FrameworkName\ can increase the overall performance by 43.86 and 62.81 percentage points.
Selective simulations utilising LLM agents playing the roles of a human expert and AI collaborators further illustrate how collaborative exploration can improve decision-making, achieving a detection success rate of up to 85.7\% on unseen intrusions in the \KDDCup\ dataset. These outcomes underscore the potential of \FrameworkName~in addressing dynamic and complex decision-making challenges across various domains.

\vspace{-2pt}
\section{Background and Related Works}
\label{sec:related}

Remarkable progress has been made in the field of machine learning (ML) with models achieving high accuracy levels in various domains~\cite{DL_survey}. However, a key limitation lies in these models' inability to recognise their knowledge boundaries~\cite{RejectionLearning_Survey_new}, resulting in overconfident errors. 

Rejection learning~\cite{RejectionLearning_Survey_old} is a concept that has been proposed to address this limitation. It enables ML models to refrain from making predictions when the risk of error is high. It involves equipping ML models with a \textit{rejector} option, enabling them to withhold predictions when faced with uncertain inputs or those outside the boundaries of their training data. 
Three common approaches exist for integrating a rejector option in ML models. In the \textit{Separated Rejector}, the rejector and predictor components remain separate, with no information exchange between them, making it modular and easy to integrate with existing systems. In the \textit{Dependent Rejector} approach, the rejector relies on the predictor's output to assess the confidence or reliability of the prediction. This approach introduces tighter coupling between the predictor and rejector components, leading to potentially better performance but limiting system flexibility. The third approach is the \textit{Integrated Rejector}, in which the rejector and predictor are merged into a single model, enabling direct interaction and joint optimisation. By sharing internal representations, the integrated rejector aims to achieve improved performance and accuracy. However, this design choice may introduce complexity and hinder interpretability. Each approach has its pros and cons, and the choice hinges on the specific problem's requirements and constraints. To ensure modular design and seamless integration with existing systems, we opted for a separated rejector design in our work (cf~\autoref{fig:Pipeline}). 

Learning to defer (L2D)~\cite{madras2018predict} builds on the rejection learning concept by combining an automated model and an external (human) expert, allowing the former to defer certain complex or uncertain decisions to the latter. L2D may include adaptive deferral, enabling the AI model to make the decision even when uncertain, particularly in cases of severe inaccuracy or bias in the human expert's judgement. Keswani et al.~\cite{keswani2021towards} extend this approach to settings where multiple experts are available. 

De et al.\cite{de2020regression} explored flexible automation levels where decisions are divided between humans and AI. They developed AI models capable of deciding which samples should be outsourced to humans based on the underlying relationship between a given sample and its corresponding human and AI model error. 
Similarly, Raghu et al.~\cite{raghu2019algorithmic} also investigated the problem of optimising the allocation of human and algorithmic effort. They trained an AI model that provides triage judgements about which instances should be handled algorithmically, and  which should be handled by human efforts. Through experiments, they demonstrated that the optimal approach is not complete automation, but rather a combination of automation and human effort.

Human-AI ensembles~\cite{choudhary2023human} represents an approach to decision-making that enhances overall accuracy by combining the results from multiple decision-makers working on the same problem, capitalising on error cancellation. In the domain of cancer detection within lymph node cell images, Wang et al~\cite{wang2016deep} demonstrated that integrating inputs from both AI and pathologists substantially improved accuracy and reduced error rates when compared to AI-exclusive or pathologist-exclusive approaches. Building upon this, Singh, Jain and Jha~\cite{singh2023subset} recently extended the concept to involve multiple humans instead of one. Through empirical evaluation, they demonstrated that combining predicted labels from multiple human experts with the probabilistic output of the AI model outperforms the AI model or any individual human in terms of accuracy. 

However, to the best of our knowledge, none of these existing works, as well as other related studies, have explicitly considered the scenario where neither the human expert nor the AI model is capable of making accurate or reliable decisions. This particular scenario poses a unique challenge that requires a more collaborative approach to joint decision-making. Our proposed framework aims to fill this gap as discussed in \autoref{sec:method}. 

\vspace{-5pt}
\section{Problem Formulation}
\label{sec:problem_formulation}

\begin{figure}[t]
    \centering
    \includegraphics[clip, trim= 27pt 29pt 27pt 29pt, width=1\linewidth]{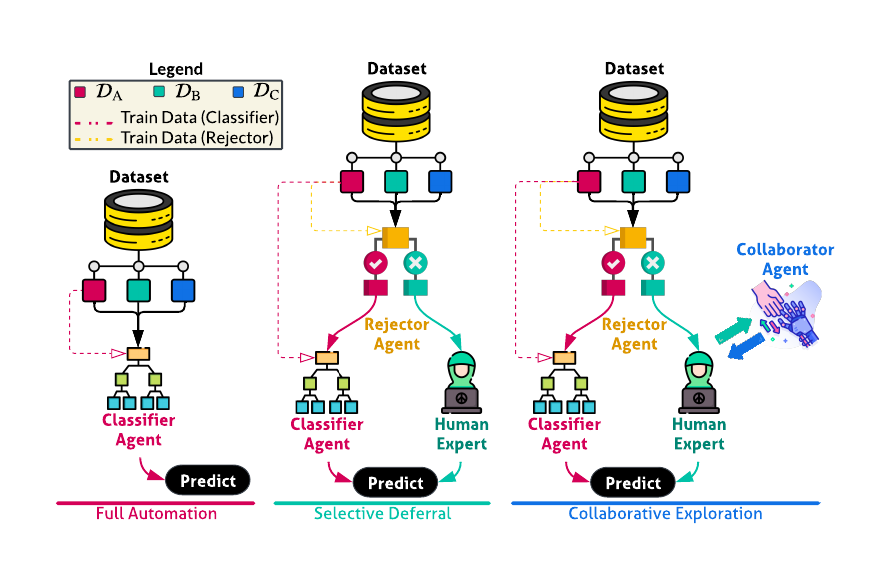}\vspace{-5pt}
    \caption{Three forms of decision-making in \textit{\FrameworkName}: \textit{full automation}, \textit{selective deferral}, and \textit{collaborative exploration}. 
    }\vspace{-15pt}
    \label{fig:Pipeline}
\end{figure}

\noindent\textbf{\textsc{Motivating Scenarios.}} 
We present a series of handcrafted examples from the cybersecurity domain to illustrate the different scenarios that our framework handles. We begin with a simple scenario and gradually introduce complexity in subsequent scenarios. To categorise these scenarios, we leverage the concepts of automation, augmentation, and collaboration. We illustrate these scenarios in \autoref{fig:Pipeline}. 

\begin{itemize}[leftmargin=*]
\item \textbf{\textsc{Full Automation.}} In the \textit{first} scenario, an anomaly and intrusion detection system (AIDS) deployed within the network infrastructure detects suspicious activity. Once detected, the system sends the corresponding network logs to the \textit{classifier}, which is trained and configured to automatically identify and respond to known threats based on the received logs.

\item \textbf{\textsc{Selective Deferral.}} In the \textit{second} scenario, the AIDS sends the network logs corresponding to the suspicious activity to the \textit{rejector}, which correlates them against predefined rules and learned patterns constituting its decision boundary. Two decision outcomes are possible: 
\begin{itemize}[leftmargin=*]
    \item If a matching rule is found, the rejector forwards the logs to the \textit{classifier}, which identifies the threat and responds accordingly (\textbf{\textsc{Full Automation}}). 
    \item Alternatively, the \textit{rejector} defers the decision to the \textit{human expert}. In this process, the rejector may offer relevant contextual information to the expert (\textbf{\textsc{Augmentation}}). The expert carefully reviews this context, in conjunction with any additional side information, such as cyber threat intelligence (CTI) reports, to arrive at an informed decision regarding how to address the detected threat.
\end{itemize}

\item \textbf{\textsc{Collaborative Exploration (CoEx)}.} In the \textit{third} scenario, the suspicious network activity logs are forwarded to the rejector, which fails to find a matching rule. This results in a deferral to the expert with relevant contextual information (\textbf{\textsc{Augmentation}}). Despite a thorough examination, the expert is unable to reach a conclusive decision based on the provided information. Recognising the need for additional information, the expert initiates collaborative exploration with a \textit{collaborator}. The collaborative effort involves an interactive and more in-depth analysis of the threat logs and consultation of CTI reports by the expert and the colaborator to uncover additional evidence. Once the expert is satisfied with the evidence obtained through this collaboration, the process is terminated, and the final decision on how to handle the threat is reached.
\end{itemize}

These three scenarios illustrate the escalating levels of complexity and uncertainty, highlighting the role of automation, augmentation, and collaboration within the decision-making process. They underscore several key points: (i) automation-based approaches excel in effectively managing well-defined and known scenarios, (ii) augmentation-based methods involve the transfer of context-rich information to the expert, enhancing their decision-making when uncertainty arises, and (iii) collaboration-based methods involve dynamic, interactive two-way information flow between experts and AI, enabling cooperative exploration, especially in situations requiring further analysis. These scenarios highlight the importance of customising the level of human-AI interaction to the specific context and requirements of each situation. 


\vspace{5pt}
\noindent\textbf{\textsc{Managing Classification Uncertainty.}}
The task of learning to reject poses a greater challenge compared to traditional classification problems. This distinction is rooted in the underlying complexities associated with handling instances that are deemed unclassifiable or uncertain. Let us consider a labelled dataset $L=\left  \{ x_i, y_i \right \}_{i=1}^n$, where each sample $x_i$ is associated with a corresponding label $y_i$ from a discrete set $y_i \in \{1, 2, \ldots, \mathcal{C}\}$, where $\mathcal{C}$ represents the total number of classes. Traditionally, the objective of classification is to learn a mapping function $f$ that maps instances from the sample space to the label space, i.e., $y'_i = f(x_i)$.

To quantify the effectiveness of the classification model $f$, we rely on a designated loss function $\mathcal{L}(y_i, y'_i)$ that measures the discrepancy between the true label $y_i$ and the predicted label $y'_i$. The classification model $f$ can be trained effectively by minimising the empirical loss over the entire training dataset, expressed as the optimisation problem $\min_f \sum_{i=1}^n \mathcal{L}(y_i, y'_i)$. This problem is well-defined and has garnered significant attention in prior research~\cite{DLinCybersecurtiy_Survey}. It is important to emphasise that the classification function $f$ primarily focuses on achieving accuracy, where the predicted label $y'_i$ closely aligns with the ground truth $y_i$ such that $\mathcal{P}(y'_i = y_i) \rightarrow 1$. However, it is crucial to acknowledge that even with a highly accurate model $f$, prediction errors are inevitable, especially when encountering data samples $x_i$ that significantly differ from the training data and belong to previously unseen classes. In such cases, the classifier lacking self-awareness would attempt to assign these samples to known classes rather than rejecting them, resulting in overconfident errors. This motivates the use of a rejector mechanism to filter out such samples.

\begin{table}[!t]
\centering
\caption{Notation and Description}\vspace{-13pt}
\label{tab:notations}
\resizebox{0.9\linewidth}{!}{%
\begin{tabular}{ll}
\toprule
\textbf{Notation} & \textbf{Description} \\
\hline
$\mathcal{D}$, $\mathcal{C}$ & Set of all data samples and set of all class labels\\
$\mathcal{D}_\textsc{a}$, $\mathcal{C}_\textsc{a}$  & Subsets of $\mathcal{D}$ and  $\mathcal{C}$ with positive samples and classes \\
$\mathcal{D}_\textsc{b}$, $\mathcal{D}_\textsc{c}$, $\mathcal{C}_\textsc{b}$, $\mathcal{C}_\textsc{c}$ & Subsets of $\mathcal{D}$ and  $\mathcal{C}$ with negative samples and classes \\
$\mathcal{D}_\textsc{k}^a$, $\mathcal{C}_\textsc{k}^a$ & Dataset and classes known to AI \\
$\mathcal{D}_\textsc{u}^a$, $\mathcal{C}_\textsc{u}^a$ & Dataset and classes unknown to AI \\
$\mathcal{D}_\textsc{k}^e$, $\mathcal{C}_\textsc{k}^e$ & Dataset and classes known to expert \\
$\mathcal{D}_\textsc{u}^e$, $\mathcal{C}_\textsc{u}^e$ & Dataset and classes unknown to expert \\
$f_r$, $f_c$, $f_e$ & Rejector, Classifier and Expert \\
\bottomrule
\end{tabular}}\vspace{-15pt}
\end{table}

\vspace{5pt}
\noindent\textbf{\textsc{Data Representation.}}
%
In order to rigorously define and subsequently analyse our proposed \FrameworkName\ framework, a systematic data representation that encapsulates the delineated scenarios is imperative. Let $\mathcal{D}$ signify the complete dataset. We partition $\mathcal{D}$ into three non-overlapping subsets: $\mathcal{D}_\textsc{a}$, $\mathcal{D}_\textsc{b}$, and $\mathcal{D}_\textsc{c}$. For each instance $x_i \in \mathcal{D}$, there exists a corresponding label $y_i \in \{1, 2, \ldots, \mathcal{C}\}$, where $\mathcal{C}$ represents the total class count.
Distinctively, each of the subsets, $\mathcal{D}_\textsc{a}$, $\mathcal{D}_\textsc{b}$, and $\mathcal{D}_\textsc{c}$, is characterised by a unique class set: $\mathcal{C}_\textsc{a}$, $\mathcal{C}_\textsc{b}$, and $\mathcal{C}_\textsc{c}$, respectively, ensuring that $\mathcal{C}_\textsc{a} \cap \mathcal{C}_\textsc{b} \cap \mathcal{C}_\textsc{c} = \varnothing$.
For AI models, the known set $\mathcal{D}_\textsc{k}^a$ consists of ${\mathcal{D}_\textsc{a}}$ while the unknown set $\mathcal{D}_\textsc{u}^a$ amalgamates ${\mathcal{D}_\textsc{b} \cup \mathcal{D}_\textsc{c}}$. Analogously, their class counterparts are $\mathcal{C}_\textsc{k}^a = {\mathcal{C}_\textsc{a}}$ and $\mathcal{C}_\textsc{u}^a = {\mathcal{C}_\textsc{b} \cup \mathcal{C}_\textsc{c}}$. Expert-defined sets depend on competency levels, detailed in \autoref{sec:experiment}.


To concretise this abstraction, consider the MNIST dataset~\cite{MNIST}. Within our schema, even digits ($\{0,2,4,6,8\}$) constitute $\mathcal{C}_\textsc{a}$, whereas odd digits are apportioned into $\mathcal{C}_\textsc{b}=\{1,3,5\}$ and $\mathcal{C}_\textsc{c}=\{7,9\}$, leading to $\mathcal{C}_\textsc{k}^a =\{0,2,4,6,8\}$ and $\mathcal{C}_\textsc{u}^a =\{1,3,5,7,9\}$. Associated notations are enumerated in \autoref{tab:notations}. This data stratification equips us with a robust basis for the nuanced development of our \FrameworkName\ framework. The well-defined divisions, capturing varying data granularities, inform tailored algorithmic solutions, ensuring the efficacy of the framework across various domains.


\section{\FrameworkName\ Framework}
\label{sec:method}
We introduce a comprehensive framework encompassing three stages, specifically designed to accommodate the increasing levels of uncertainty associated with the data used for decision-making. Our framework integrates multiple AI components, namely \textit{rejector, classifier,} and \textit{collaborator}, in conjunction with the \textit{human expert}. In this section, we outline the sequential flow of data and decision-making within our framework, emphasising the distinct roles of each component and their collaborative interactions. This sequential flow showcases the modular design and collaborative nature of our approach, facilitating effective decision-making in complex and uncertain scenarios.

\vspace{5pt}
\noindent \textbf{\textsc{Rejector}. } 
Within our framework, each data sample $x_i$ is first provided to a rejector, whose primary function is to determine the appropriate forwarding destination. Samples belonging to $\mathcal{D}_\textsc{k}^a$ are forwarded to the classifier, while the remaining samples are deferred to the human expert for further analysis and decision-making. The learning to reject task can be viewed as a binary classification problem, distinguishing between acceptance (positive) and rejection (negative), i.e., $\mathcal{D}_\textsc{k}^a$  and $\mathcal{D}_\textsc{u}^a$, respectively. However, this presents a challenge as negative samples $\mathcal{D}_\textsc{u}^a$, indicative of rejected instances, are typically unavailable, rendering it an unbalanced one-class problem. Given that the only data available to the rejector for training is the training set from $\mathcal{D}_\textsc{k}^a$, we can define the rejector's decision-making function $f_r$ as a one-class classifier. Let $X_k = \{x_i \in \mathcal{D}_\textsc{k}^a\}$ represent the set of data samples belonging to $\mathcal{D}_\textsc{k}^a$, which serves as the positive training set for the rejector. The objective of $f_r$ is to learn a decision boundary that effectively separates the positive samples in $X_k$ from the unobserved negative samples in $\mathcal{D}_\textsc{u}^a$. 

Given a data sample $x_i$, the rejector computes a score or a measure of \textit{acceptance} based on its learned model. This score, denoted as $s(x_i)$, quantifies the degree of compatibility between $x_i$ and the positive class. To determine whether $x_i$ should be forwarded to the classifier or deferred to the expert, a decision threshold $\theta_r$ is introduced. If $s(x_i)$ exceeds $\theta_r$, $x_i$ is classified as belonging to $\mathcal{D}_\textsc{k}^a$ and is forwarded to the classifier. Otherwise, if $s(x_i)$ is below or equal to $\theta_r$, $x_i$ is considered as a potential negative sample and is deferred to the expert for further analysis. The formulation of $f_r(x_i)$ can be summarised as follows:
\begin{equation}
\small
f_r(x_i)=
    \begin{cases}
1, & \text{if } s(x_i) > \theta_r \\
0, & \text{otherwise}
\end{cases}
\end{equation}
Here, $f_r(x_i)$ denotes the output of the rejector $f_r$ for a given data sample $x_i$, where 1 indicates acceptance and 0 indicates rejection. The decision threshold $\theta_r$ is determined based on the learned model and serves as a criterion for the rejector to make decisions on the data samples. We discuss different one-class model architecture choices for the rejector in the \autoref{sec:experiment}.

\vspace{5pt}
\noindent\textbf{\textsc{Classifier}. } When the data sample $x_i$ belongs to the known set $\mathcal{D}_\textsc{k}^a$, the rejector forwards it to the classifier $f_c$, which is responsible for assigning a class label to the sample $x_i$ from the known classes $\mathcal{C}_\textsc{k}^a$.
To address this scenario, we adopt a traditional classification approach, aiming to learn a mapping function $f_c$ that accurately predicts the labels $y_i^k$ given the corresponding input examples $x_i^k$. This can be formulated as an optimisation problem:
\begin{equation}
\small
\min_{f_c} \sum_{i=1}^n \mathcal{L} (y_i^k, \hat{y}_i^k) \quad \text{where} \quad \hat{y}_i^k = f_c(x_i^k)
\end{equation}
Here, $\mathcal{L}$ represents an appropriate loss function, such as cross-entropy or mean squared error, quantifying the discrepancy between the predicted labels $\hat{y}_i^k$ and the true labels $y_i^k$. The objective is to minimise this loss over the training dataset, which consists of pairs of input examples $x_i^k$ and their corresponding true labels $y_i^k$. By minimising this objective function, we enable the automation of the entire process, allowing the classifier to effectively handle all data samples belonging to $\mathcal{C}_\textsc{k}^a$. This automation reduces the workload on experts, as $f_c$ can accurately predict the labels for known data samples.

\vspace{5pt}
\noindent\textbf{\textsc{Expert.} }
When the data sample $x_i$ does not belong to the known set $\mathcal{D}_\textsc{k}^a$, i.e., $x_i \in \mathcal{D}_\textsc{u}^a$, the rejector defers it to the expert $f_e$. During this stage, the expert is provided with contextual information regarding the reasons for sample rejections by $f_r$. Furthermore, the expert is presented with the prediction probabilities associated with the data sample if it were to be passed to the classifier, along with additional context derived from side information such as CTI reports. Using this information, the expert can make informed decisions and assign an appropriate label to sample $x_i$.

Let us denote the expert's decision for $x_i$ as $f_e(x_i)$, which represents the label assigned to $x_i$ by the expert. When $x_i$ is deferred, the expert is provided with contextual information, denoted as $I_c(x_i)$, that explains the reasons for the $f_r$ decision to defer $x_i$. Additionally, the expert is presented with the prediction probabilities associated with $x_i$ if it were passed to the $f_c$, denoted as $P({y'}_i | x_i)$, where ${y'}_i$ is the predicted label for $x_i$ by $f_c$. Furthermore, supplementary context derived from side information $I_s(x_i)$, such as CTI reports, is also available to the expert. Equipped with this information, the expert can make informed decisions about $x_i$ and either assign it a label, represented as $f_e(x_i) = \text{label}$, or engage with the collaborator. We can represent this concept using the following equation:
\begin{equation}
\small
    f_e(x_i)=\mathrm{ExpertDecision} (x_i, f_r(x_i), P({y'}_i | x_i), I_c(x_i), I_s(x_i))
\end{equation}

\vspace{5pt}
\noindent\textbf{\textsc{Collaborator}. }
If the expert is unable to reach a conclusive decision regarding the data sample $x_i$, it engages with the collaborator to jointly perform exploratory investigations to try and reach a final decision. This collaborative process resembles the partnership between two human experts engaged in cyber threat analysis. In our framework, the expert has the option to instruct the collaborator to conduct further analyses, while the AI system can suggest actions conducive to human decision-making. This iterative process continues until a final decision is reached, combining the strengths of both humans and AI and enhancing the system's overall effectiveness.

During the exploratory collaboration stage, we have the expert ($E$) and the collaborator ($A$) as the decision-makers. We represent this collaboration using a Bayesian network, where nodes represent entities and edges represent information flow. Initially, both $E$ and $A$ have prior beliefs ($P(E)$ and $P(A)$) reflecting their initial knowledge and subjective opinions. As the collaboration unfolds, these nodes update their beliefs based on observed evidence ($O$) and shared information, using Bayes' rule. The expert's updated belief $P(E | O)$ is calculated as $\frac{P(O | E) \cdot P(E)}{P(O)}$. Similarly, the collaborator's updated belief $P(A | O)$ is computed as $\frac{P(O | A) \cdot P(A)}{P(O)}$. The nodes can communicate through prompts, requests for analysis, or sharing of results, facilitating a back-and-forth interaction. The collaboration iterates as the nodes exchange information and refine their beliefs until they reach a consensus or the expert makes a decision.

\section{Experimental Settings}
\label{sec:experiment}
In this section, we outline the experimental setup for assessing the performance of each component within our proposed \FrameworkName\ framework, addressing the following key research questions: 

\begin{itemize}[leftmargin=*]
    \item \textbf{RQ1.} What is the performance of the \FrameworkName~framework when automated classification is the sole decision-making component? 
    \item \textbf{RQ2.} How does the performance of the \FrameworkName~framework change when automated classification is combined with selective deferral to a human expert? 
    \item \textbf{RQ3.} To what extent does the integration of collaborative exploration enhance the performance of the \FrameworkName~framework?
\end{itemize}

By tackling these questions, we aim to provide a comprehensive evaluation of the system's performance under different conditions, shedding light on the limitations of full automation, the role of the rejector, and the advantages of CoEx within the proposed framework. We also investigate the performance of different rejector architectures applied to the chosen datasets. This allows us to identify the most effective rejection function for subsequent experiments. 

\begin{table}
\centering
\caption{Class Assignments}\vspace{-13pt}
\label{tab:class-assignments}
\resizebox{0.9\linewidth}{!}{%
\begin{tabular}{l|l}
\toprule
\textbf{\textit{Dataset}} & \textbf{\textit{Class Assignments}} \\ 
\hline
\MNIST & $\mathcal{C}_\textsc{a} = \{0,2,4,6,8\}$, $\mathcal{C}_\textsc{b} = \{1,3,5\}$,  $\mathcal{C}_\textsc{c} = \{7,9\}$ \\ 
\hline
\fMNIST & $\mathcal{C}_\textsc{a} = \{$``T-Shirt/Top'', ``Pullover'', ``Coat'', ``Shirt'', ``Bag''$\}$ \\
 & $\mathcal{C}_\textsc{b} = \{$``Trouser'', ``Dress'', ``Sandals''$\}$, $\mathcal{C}_\textsc{c} = \{$``Sneaker'', ``AnkleBoots''$\}$ \\

\hline
\CIFAR-10 & $\mathcal{C}_\textsc{a} = \{$``Airplane'', ``Bird'', ``Deer'', ``Frog'', ``Ship''$\}$ \\
 & $\mathcal{C}_\textsc{b} = \{$``Automobile'', ``Cat'', ``Dog''$\}$, $\mathcal{C}_\textsc{c} = \{$``Horse'', ``Truck''$\}$ \\ 
\hline
\KDDCup & $\mathcal{C}_\textsc{a} = \{$``Back'', ``Land'', ``Pod'', ``Smurf'', ``Teardrop''$\}$ \\
 & $\mathcal{C}_\textsc{b} = \{$``BufferOverview'', ``FtpWrite'', ``GuessPasswd'',``IMap'', ``IpSweep'', ``Perl'', \\ & ``PortSweep'', ``RootKit'', ``Satan'', ``WarezClient''$\}$, $\mathcal{C}_\textsc{c} = \{$``LoadModule'',\\& ``Multihop'', ``Neptune'', ``NMap'', ``Phf'', ``Spy'', ``WarezMaster''$\}$\\
 \bottomrule
\end{tabular}
}\vspace{-10pt}
\end{table}

\vspace{5pt}
\noindent\textbf{\textsc{Datasets}. }
To rigorously assess the \FrameworkName\ framework, we used four popular benchmark datasets: \MNIST~\cite{MNIST}, \fMNIST~\cite{fMNIST}, \CIFAR-10~\cite{CIFAR}, and \KDDCup~\cite{KDDCup99}. We chose these datasets because they have been commonly used in prior research~\cite{DeepMAD,UniAD,FCDD+OE,PANDA}, particularly in one-class settings, mirroring our problem setting. As illustrative instances, Singh et al.\cite{DeepMAD} and You et al.\cite{UniAD} both employed analogous divisions on these datasets, harmonizing with our classification of $\mathcal{C}_\textsc{k}^a$ and $\mathcal{C}_\textsc{u}^a$. \KDDCup, a network traffic dataset, further bolsters our experiment, elucidating the framework's suitability for cybersecurity contexts.


Each dataset, including \KDDCup, underwent meticulous partitioning to accommodate evaluations across scenarios delineated in~\autoref{sec:problem_formulation}. Distinctively, we demarcated subsets: $\mathcal{D}_\textsc{a}$, $\mathcal{D}_\textsc{b}$, and $\mathcal{D}_\textsc{c}$, rooted in known ($\mathcal{C}_\textsc{a}$) and unknown classes ($\mathcal{C}_\textsc{b}$ and $\mathcal{C}_\textsc{c}$)\footnote{Within our \KDDCup\ context, the normal class modelling is not requisite since only anomalies, per motivating scenarios in \autoref{sec:problem_formulation}, are relayed to the rejector, concentrating the analysis solely on anomaly handling.}. The specific class assignments for these subsets are elaborated in \autoref{tab:class-assignments}.

\textit{Datasets Configuration. } 
In this section, we highlight the specific configurations used for each dataset. For \MNIST\ and \fMNIST, we employed $6000\times5$ samples in $\mathcal{D}_\textsc{a}$, $1000\times3$ samples in $\mathcal{D}_\textsc{b}$, and $1000\times2$ samples in $\mathcal{D}_\textsc{c}$. The \CIFAR-10 dataset was structured with $5000\times5$ samples in $\mathcal{D}_\textsc{a}$, $1000\times3$ samples in $\mathcal{D}_\textsc{b}$, and $1000\times2$ samples in $\mathcal{D}_\textsc{c}$. 
However, the \KDDCup\ dataset posed a unique challenge due to its varying class sample distribution. To address this, we randomly allocated the classes into $\mathcal{D}_\textsc{a}$, $\mathcal{D}_\textsc{b}$, and $\mathcal{D}_\textsc{c}$, resulting in class sample counts of $284257$, $5012$, and $107474$ respectively (see \autoref{tab:class-assignments}). 
Notably, during the training of classifiers and rejectors, only positive class samples (i.e., $\mathcal{D}_\textsc{a}$) were used, and we adopted an 80-20 split strategy. 

\begin{table}[t]
\centering
\caption{Rejector baseline for different datasets.}\vspace{-13pt}
\label{tab:baselines}
\resizebox{0.9\linewidth}{!}{%
\begin{tabular}{l|l} 
\toprule
\textbf{\textit{Dataset}} & \textbf{\textit{Baselines}} \\ 
\hline
\MNIST & \begin{tabular}[c]{@{}l@{}}OCSVM~\cite{OCSVM}, Deep Autoencoder (DeepAE), \\Convolutional AE (ConvAE), and DeepSVDD~\cite{DeepSVDD}\end{tabular} \\ 
\hline
\fMNIST & \begin{tabular}[c]{@{}l@{}}OCSVM, DeepMAD~\cite{DeepMAD}, DROCC~\cite{DROCC}, and DeepSVDD~\cite{DeepSVDD}\end{tabular} \\ 
\hline
\CIFAR-10 & \begin{tabular}[c]{@{}l@{}}PANDA~\cite{PANDA}, DeepMAD, FCDD+OE\cite{FCDD+OE}, UniAD~\cite{UniAD},  \\US~\cite{US}, DeepSVDD~\cite{DeepSVDD}, DROCC, and MKD~\cite{MKD}\end{tabular} \\ 
\hline
\KDDCup & OCSVM~\cite{OCSVM}, DAGMM~\cite{DAGMM}, OGNet~\cite{OGNet}, and MemAE~\cite{MemAE}\\
\bottomrule
\end{tabular}
}\vspace{-10pt}
\end{table}

\vspace{5pt}
\noindent\textbf{\textsc{Rejector Settings}. }
The rejector selectively directs samples to the classifier based on their conformance to the classifier's training data and defers the rest to the expert. 
In this context, the utilisation of a conventional two-class classification solution is rendered infeasible due to the absence of negative samples $\mathcal{D}_\textsc{u}^a$. Therefore, we rely on methods that solely utilise the positive class, denoted as $\mathcal{D}_\textsc{k}^a$, to establish a discerning decision boundary. One-class methods are a prominent approach for accomplishing this objective. Hence, we have selected the best-performing one-class methods identified in recent literature to serve as the rejector in our framework. We present the methods used for evaluation in \autoref{tab:baselines}. 

We evaluate the rejector across two distinct settings. Firstly, we assess the inherent capabilities of various rejector candidates in a one-class setting across multiple datasets, aiming to quantify their raw performance. Secondly, we employ the rejector within our framework to evaluate the \textit{selective deferral} and \textit{collaborative exploration} scenarios.

\vspace{5pt}
\noindent\textbf{\textsc{Classifier Settings}. }
Compared to the rejector, the classifier's task is relatively straightforward, involving knowledge acquisition from the training data $\mathcal{D}_\textsc{k}^a$ and subsequently classifying the test data. Thus, the classifier task can be regarded as a conventional multiclass classification problem. We employed established methods for this task, including Support vector machines (SVMs)~\cite{SVM} and Random Forests (RF)~\cite{RF}, (FFNs)~\cite{FFNs} (with 3 linear layers), and convolutional neural networks (CNNs)~\cite{AlexNet} (with 2 convolutional layers and 2 linear layers) to assess the classifier's performance. We deliberately opted for simpler methods because the classification task is relatively uncomplicated. Nevertheless, \FrameworkName~allows easy integration of more advanced approaches like generative adversarial networks (GANs)~\cite{GANs} or transformers~\cite{TransformersSurvey} if needed. 

We evaluate the classifier in two unique settings. In the full automation mode, the classifier operates as the primary decision-making component within the system, assuming complete control over the classification process. In the second setting, the classifier functions as an integral part of the broader framework, wherein we specifically explore \textit{selective deferral} and \textit{collaborative exploration}. 


\vspace{5pt}
\noindent\textbf{\textsc{Expert Settings}. }
In scenarios where the rejector encounters a data sample $x_i$ that does not belong to the positive class, it opts to defer to the expert. Our evaluation encompasses different levels of decision-making competence, denoted by the variable $t$, which reflects the expert's proficiency in identifying data samples from a specific class. We define three distinct competence settings as follows.
The first competence setting ($t=1$) assumes that the expert possesses the ability to accurately recognise the positive classes   $\mathcal{C}_\textsc{k}^e = \mathcal{C}_\textsc{a} = \mathcal{C}_\textsc{k}^a$  (i.e., the same classes used to train rejector and classifier)\footnote{Deferring samples to the expert does not necessarily ensure complete accuracy. The expert may still encounter false positives in their analysis. To address this possibility, we have set $t=1$ in our experiments. Additionally, the classifier's performance inherently accounts for false negatives through a proportional reduction.}.
The second competence setting ($t=2$) assumes that the expert possesses the capability to recognise all the classes in $\mathcal{C}_\textsc{b}$ (i.e., $\mathcal{C}_\textsc{k}^e = \mathcal{C}_\textsc{b}$) by using side information and contextual data. Finally, the third competence setting ($t=3$) assumes that the expert can identify both positive classes ($\mathcal{C}_\textsc{a}$) as well as a subset of negative classes ($\mathcal{C}_\textsc{b}$), i.e., $\mathcal{C}_\textsc{k}^e = \{\mathcal{C}_\textsc{a} \cup \mathcal{C}_\textsc{b}\}$. By encompassing a broader range of competencies in these settings, we aim to comprehensively investigate the impact of varying human decision-making abilities on the overall classification process. It is important to note that prior research has employed analogous methodologies to emulate human competence, particularly within the domain of L2D~\cite{L2D_ConsistentEstimators}. 

We undertake an evaluation to assess the varying levels of expertise exhibited by the expert in both \textit{selective deferral} and \textit{collaborative exploration} scenarios.

\vspace{5pt}
\noindent
\textbf{\textsc{Collaborator Settings}. }
When the expert encounters a data sample $x_i \in \mathcal{D}_\textsc{c}$ from the rejector, arriving at a definitive decision is challenging, even when considering the available side information and contextual data. In other words, both the classifier and the expert are uncertain in this case. To address this limitation of \textit{selective deferral} scenario, our \FrameworkName\ framework offers the expert an opportunity to engage in a collaborative exploration (CoEx) with the collaborator, aiming to acquire supplementary information concerning $x_i$ that may facilitate the expert in reaching a resolution. However, it is pivotal to acknowledge the subjectivity inherent in this step; its effectiveness is contingent upon various factors and might differ across individuals. 
To investigate the potential outcomes of this collaborative exploration, we examine four distinct scenarios: $r=1$, $r=2$, $r=3$, and $r=4$, representing resolution rates of 50\%, 75\%, 90\%, and 100\%, respectively, on \MNIST, \fMNIST, \CIFAR-10, and \KDDCup~datasets.

\vspace{5pt}
\noindent
\textbf{\textsc{Simulation of CoEx using Virtual Personas}.}
To further validate CoEx, we design experiments leveraging virtual personas.

\textit{Persona Design:} We create three distinct virtual personas representing security analysts with varying experience levels: Jordan (novice), Alex (intermediate), and John (expert). Additionally, we develop a collaborator persona, dubbed SentinelBot, endowed with specific competencies, including advanced data analytics\footnote{More details regarding personas: \url{https://anonymous.4open.science/r/A2C/personas}}. The design of our personas is influenced by recent advancements in generative agents, particularly the work by Park et al.~\cite{park2023generative}. However, our personas incorporate a significantly richer set of details. 

\textit{Dataset and Initial Evaluation:} The simulation utilises \KDDCup\ dataset, focusing predominantly on intrusion detection---a domain aligned with the designated expertise of the security analyst personas. Our preliminary evaluations (benchmarking) considered known segments of the dataset, categorised into four distinct sample types: Normal (\textit{Norm.}), Rare Normal (\textit{Norm. (Rare)}), Intrusion (\textit{Intru.}), and Rare Intrusion (\textit{Intru. (Rare)}). SentinelBot's performance in isolation served as our baseline metric.

\textit{CoEx Assessment:} Following the baseline evaluations, collaborative explorations are initiated, pairing SentinelBot with each security analyst persona, yielding configurations such as $SentinelBot + Jordan$, among others. These collaborative configurations are subjected to evaluations under three distinct competence tiers, represented as $t=1$, $t=2$, and $t=3$. Within this setup, for any given sample $x_i \in \mathcal{D}_\textsc{c}$, the analyst persona engages in iterative exchanges with SentinelBot, leveraging historical data defined by the competence tier as context. This dialogue converges once the security analyst determines a conclusive decision for $x_i$, with potential outcomes being \textit{normal}, \textit{intrusion}, or a \textit{call for caution}.

\textit{Decision Markings:} Correct identifications of intrusion (or normal) samples are represented with $\checkmark$. Incorrect ones are indicated by the $\times$ mark. Situations warranting caution are marked with \raisebox{.40ex}{{\fontencoding{U}\fontfamily{futs}\selectfont\char 49 \relax}}. 

\textit{Implementation:} The interactive dialogues between the personas and SentinelBot are orchestrated using distinct instances of the GPT-4 model from ChatGPT. Notably, SentinelBot is endowed with an enhanced capability for sophisticated data analysis. Comprehensive insights into the conversation dynamics and other intricate facets of the implementation can be found in the accompanying source code\footnote{The anonymised code for the LLM-agents based simulation is accessible at \textcolor{teal}{\url{https://anonymous.4open.science/r/A2C}.} We will release the public version post-acceptance.}. For the sake of experimental pragmatism and to ensure a comprehensive representation, we curate one sample within \(\mathcal{D}_\textsc{c}\) corresponding to each of the seven intrusion categories. This results in a total of $75$ CoEx experiments. It is important to note that conducting a full-scale experiment on the entire \(\mathcal{D}_\textsc{c}\), encompassing 107,474 samples, would be computationally challenging and unfeasible within our current setup. We discuss this further in \autoref{sec:discussion}. \textit{Note: We provide excerpts of some interesting conversations between the personas and SentinelBot in the \autoref{app:persona_conversation}. }

\vspace{5pt}
\noindent
\textbf{\textsc{Performance Metrics}. } We use the F1-score, a prevalent and trusted measure for classification tasks, as our primary metric of choice. Furthermore, for our persona-based experiment, we introduce the collaborative exploration success rate metric $\textsc{CoEx}_\textsc{sr}$, calculated as: $\textsc{CoEx}_\textsc{sr} = \frac{\sum_{i=1}^{n} s_{i}}{I \times 100}$. Here, $s_{i}$ denotes the score attributed to the $i^{th}$ prediction, taking values 0, 0.5, or 1, depending on the decision outcomes symbolised by $\times$, \raisebox{.40ex}{{\fontencoding{U}\fontfamily{futs}\selectfont\char 49 \relax}}, and $\checkmark$ respectively, $n$ denotes the total sample count, and $I$ represents the comprehensive count of intrusion varieties.

\vspace{5pt}
\noindent
\textbf{\textsc{Reproducibility and Consistency}. } To ensure reproducibility and transparency, we meticulously follow the specifications outlined in the code repositories of the baseline models, accounting for variations in the required libraries. This rigorous approach ensures transparent evaluations and equitable comparisons.

\section{Results}
\label{sec:results}




In this section, we present the results of our simulative experiments, with a specific focus on answering the three research questions (RQs). Each subsection provides a succinct analysis of the findings, their implications, and relevance to the particular research question. 
 
 


\noindent\textbf{\textsc{RQ1. What is the performance of the \FrameworkName~framework when automated classification is the sole decision-making component? }} 
To answer RQ1, we evaluated the efficacy of a classifier ($f_c(\cdot)$) in performing classification tasks without employing any rejector or collaborator. The classifier was trained on $\mathcal{D}_\textsc{a}$, which contains samples solely from known classes ($\mathcal{C}_\textsc{a}$). 

In \autoref{tab:standalone}, column $f_c(\mathcal{D}_\textsc{a})$ presents the classifiers' performance results on the test set of $\mathcal{D}_\textsc{a}$ while column $f_c(\mathcal{D})$ presents the performance results on the entire dataset $\mathcal{D}$, containing samples from both known and unknown classes.
The results show that all classifiers perform well on $\mathcal{D}_\textsc{a}$. As expected, deep learning methods FFN and CNN outperform classical methods, SVM and RF, achieving F1-scores of up to 99\% on certain datasets. 
%
Conversely, as shown in column $f_c(\mathcal{D})$, classifier performance drops significantly when presented with samples from unknown classes, $\mathcal{C}_\textsc{u}$. Some classifiers' performance drops by a stark 66.6\% relative to their $f_c(\mathcal{D}_\textsc{a})$ performance. 

These results highlight the limitations of relying solely on automation and motivate the use of selective deferral within the \FrameworkName\ framework to address the challenge of handling scenarios that the classifier is unfamiliar with. 

\begin{table}[t]
\centering
\caption{Performance of Standalone Classifier}
\vspace{-13pt}
 \label{tab:standalone}
\resizebox{\linewidth}{!}{%
\begin{tabular}{l|rr|rr|rr|rr} 
\toprule
\multirow{2}{*}{\textbf{Classifiers}} & \multicolumn{2}{c|}{\textbf{\MNIST}} & \multicolumn{2}{c|}{\textbf{\fMNIST}} & \multicolumn{2}{c|}{\textbf{\CIFAR-10}} & \multicolumn{2}{c}{\textbf{\KDDCup}} \\ 
\cline{2-9}
 & \multicolumn{1}{c}{$f_c(\mathcal{D}_\textsc{a})$}  & \multicolumn{1}{c|}{$f_c(\mathcal{D})$} & \multicolumn{1}{c}{$f_c(\mathcal{D}_\textsc{a})$}  & \multicolumn{1}{c|}{$f_c(\mathcal{D})$} & \multicolumn{1}{c}{$f_c(\mathcal{D}_\textsc{a})$}  & \multicolumn{1}{c|}{$f_c(\mathcal{D})$} & \multicolumn{1}{c}{$f_c(\mathcal{D}_\textsc{a})$}  & \multicolumn{1}{c}{$f_c(\mathcal{D})$} \\ 
\hline
\textbf{SVM} & 71.32\% & 35.13\% & 49.96\% & 24.98\% & 37.68\%  & 18.84\% & 99.119\% & 33.312\% \\
\textbf{RF} & 73.57\% & 36.24\% & 55.76\% & 27.88\% & 41.26\%  & 20.63\% & 99.122\% &  33.313\% \\
\textbf{FFN} & 98.66\% & 48.60\% & 84.14\% & 42.07\% & 64.94\% & 32.47\% & 99.273\% &  33.364\% \\
\textbf{CNN} & 99.45\% & 48.99\% & 87.86\% & 43.93\% & 74.76\%  & 37.38\% & 99.460\% & 33.426\% \\
\bottomrule
\end{tabular}
}\vspace{-10pt}
\end{table}

\vspace{5pt}
\noindent\textbf{\textsc{RQ2. How does the performance of the \FrameworkName~framework change when automated classification is combined with selective deferral to a human expert? }}
%
To answer RQ2, we initially conducted experiments to assess the rejector performance when using different techniques, with the aim of identifying the most suitable rejector, denoted as $f_r(\cdot)$, for the selected datasets. 
To achieve this, the rejector was trained strictly on dataset $\mathcal{D}_\textsc{a}$, encompassing exclusively positive class instances from $\mathcal{C}_\textsc{k}^a$. The rejector's competence was subsequently assessed against the complete dataset, $\mathcal{D}$. Based on this comprehensive analysis, the best-performing rejectors for each dataset were identified as follows: DeepSVDD for \MNIST\ and \fMNIST, UniAD for \CIFAR-10, and MemAE for \KDDCup. \autoref{tab:rejectors} depicts their performance results against each dataset. As expected, deep learning methodologies exhibited superior performance when compared to classical approaches. Notably, for each dataset, at least one rejector achieved performance levels surpassing 90\%, thereby effectively reducing the occurrence of false positives. 
%



Next, we conducted experiments to quantify the performance enhancement resulting from the integration of rejectors into the \FrameworkName\ framework. 
To achieve this, we employed a three-tiered competence scheme to assess deferral to experts. 
The results in \autoref{tab:collaboration}, corresponding to the three expert competence levels, $t=1$ to $t=3$, and $r=\varnothing$ (indicating no CoEx) represent the improvement in performance when automation is combined with selective deferral.

The results demonstrate a significant improvement in performance as competence levels increase. On the \MNIST\ dataset, performance surges by up to 58.03\% relative to pure automation, with similar increases of 61.07\% and 73.22\% on the \fMNIST\ and \CIFAR-10 datasets, respectively. However, the \KDDCup\ dataset presents only a nominal improvement. A closer examination of the dataset reveals that this is primarily due to its inherent composition, and the way it is split into $\mathcal{D}_\textsc{a}$, $\mathcal{D}_\textsc{b}$, and $\mathcal{D}_\textsc{c}$. The dominance of rare attacks within $\mathcal{C}_\textsc{b}$ class diminishes the sample size, thereby dampening performance impacts. In contrast, the $\mathcal{C}_\textsc{c}$ class, featuring fewer rare intrusions, manifests a prevailing class with an ample sample count. This accentuates a limitation in the deferral phase—the skewness of class distribution—motivating the subsequent collaborative exploration stage in our framework. 

\begin{table}[t]
\centering
\caption{Performance of Different Rejectors}\vspace{-13pt}
\label{tab:rejectors}
\resizebox{\linewidth}{!}{%
\begin{tabular}{lc|lc|lc|lc}
\hline
 \multicolumn{2}{c|}{\textbf{\MNIST}} & \multicolumn{2}{c|}{\textbf{\fMNIST}} & \multicolumn{2}{c|}{\textbf{\CIFAR-10}} & \multicolumn{2}{c}{\textbf{\KDDCup}} \\ \cline{1-8} 
 \textit{Rejector} & \textit{Acc.} & \textit{Rejector} & \textit{Acc.} & \textit{Rejector} & \textit{Acc.} & \textit{Rejector} & \textit{Acc.}  \\ \hline
OCSVM & 80.08\% & OCSVM & 75.36\% & PANDA & 77.10\% & OCSVM & 79.54\% \\
DeepAE & 81.06\% & DeepMAD & 88.32\% & DeepMAD & 77.38\% & DAGMM & 93.69\% \\
ConvAE & 81.06\% & DROCC & 92.60\% & FCDD+OE & 85.30\% & OGNet & 95.90\% \\
DeepSVDD & 95.36\% & DeepSVDD & 92.79\% & UniAD & 92.97\% & MemAE & 96.41\% \\ \hline
\end{tabular}%
}\vspace{-10pt}
\end{table}

\begin{table*}[t]
\centering
\caption{Performance improvement using \FrameworkName\ framework. The best-performing rejector and classifier are used for each dataset. Here, $t$ is the expert's competence, and $r$ is the problem resolution rate of Human-AI collaboration. We experimented with a resolution rate of $50\%$, $75\%$, $90\%$, and $100\%$ represented by $r=1$, $r=2$, $r=3$, and $r=4$, whereas $r=\varnothing$ represents no collaboration.}\vspace{-10pt}
\label{tab:collaboration}
\resizebox{0.9\linewidth}{!}{%
\begin{tabular}{l|rrrrr|rrrrr|rrrrr} 
\toprule
\multirow{2}{*}{\textbf{Datasets}} & \multicolumn{5}{c|}{\textbf{$t=1$}} & \multicolumn{5}{c|}{\textbf{$t=2$}} & \multicolumn{5}{c}{\textbf{$t=3$}} \\ 
\cline{2-16}
 & \multicolumn{1}{c}{$r=\varnothing$} & \multicolumn{1}{c}{$r=1$} & \multicolumn{1}{c}{$r=2$} & \multicolumn{1}{c}{$r=3$} & \multicolumn{1}{c|}{$r=4$} & \multicolumn{1}{c}{$r=\varnothing$} & \multicolumn{1}{c}{$r=1$} & \multicolumn{1}{c}{$r=2$} & \multicolumn{1}{c}{$r=3$} & \multicolumn{1}{c|}{\textbf{$r=4$}} & \multicolumn{1}{c}{\textbf{$r=\varnothing$}} & \multicolumn{1}{c}{\textbf{$r=1$}} & \multicolumn{1}{c}{\textbf{$r=2$}} & \multicolumn{1}{c}{\textbf{$r=3$}} & \multicolumn{1}{c}{\textbf{$r=4$}} \\ 
\hline
\textbf{\MNIST} & 49.74\% & 58.58\% & 63.00\% & 65.65\% & 67.42\% & 75.10\% & 83.94\% & 88.36\% & 91.01\% & 92.78\% & 77.42\% & 86.26\% & 90.68\% & 93.33\% & 95.10\% \\
\textbf{\fMNIST} & 44.37\% & 52.57\% & 56.66\% & 59.12\% & 60.76\% & 67.16\% & 75.36\% & 79.45\% & 81.91\% & 83.55\% & 70.76\% & 78.96\% & 83.06\% & 85.52\% & 87.16\% \\
\textbf{\CIFAR-10} & 38.27\% & 46.51\% & 50.63\% & 53.10\% & 54.75\% & 61.24\% & 69.48\% & 73.60\% & 76.07\% & 77.72\% & 64.75\% & 72.99\% & 77.12\% & 79.59\% & 81.24\% \\
\textbf{\KDDCup} & 34.61\% & 65.14\% & 80.40\% & 89.56\% & 95.67\% & 32.79\% & 63.32\% & 78.59\% & 87.75\% & 93.85\% & 35.18\% & 65.71\% & 80.97\% & 90.13\% & 96.24\% \\
\bottomrule
\end{tabular}
}\vspace{-5pt}
\end{table*}

\begin{table*}
\centering
\caption{Collaborative exploration using SOC analyst personas for intrusion detection in \KDDCup\ dataset. Here, ``$\checkmark$'' and ``$\times$'' denote correct and incorrect classifications, whereas ``{\WarningMark}'' represents when caution is suggested instead of a definite answer.} \vspace{-10pt}
\label{tab:persona}
\resizebox{\linewidth}{!}{%
\begin{tabular}{l|c|cccc|ccccccc|c|ccccccc|c|ccccccc|c} 
\toprule
\multicolumn{2}{c|}{\textbf{Persona}} & \multicolumn{4}{c|}{\textbf{Benchmark on known $\mathcal{D}_\textsc{a}$}} & \multicolumn{24}{c}{\textbf{Evaluation of the Personas working with SentinelBot in collaborative exploration setting to identify intrusions in unknown dataset $\mathcal{D}_\textsc{c}$}} \\ 
\hline
\multirow{2}{*}{\textit{Name}} & \multirow{2}{*}{\begin{tabular}[c]{@{}c@{}}\textit{Exp.}\\\textit{(Years)}\end{tabular}} & \multirow{2}{*}{\textit{Norm.}} & \multirow{2}{*}{\begin{tabular}[c]{@{}c@{}}\textit{Norm.}\\\textit{(Rare)}\end{tabular}} & \multirow{2}{*}{\textit{Intru.}} & \multirow{2}{*}{\begin{tabular}[c]{@{}c@{}}\textit{Intru.}\\\textit{(Rare)}\end{tabular}} & \multicolumn{8}{c|}{\textbf{Competence Setting:} $t=1$ ($\mathcal{C}_\textsc{k}^e = \mathcal{C}_\textsc{a}$)} & \multicolumn{8}{c|}{\textbf{Competence Setting:} $t=2$ ($\mathcal{C}_\textsc{k}^e = \mathcal{C}_\textsc{b}$)} & \multicolumn{8}{c}{\textbf{Competence Setting:} $t=3$ ($\mathcal{C}_\textsc{k}^e = \{\mathcal{C}_\textsc{a} \cup \mathcal{C}_\textsc{b}\}$)} \\ 
\cline{7-30}
 &  &  &  &  &  & \multicolumn{8}{c|}{\textbf{Performance on intrusion classes in }$\mathcal{D}_\textsc{c}$} & \multicolumn{8}{c|}{\textbf{Performance on intrusion classes in }$\mathcal{D}_\textsc{c}$} & \multicolumn{8}{c}{\textbf{Performance on intrusion classes in }$\mathcal{D}_\textsc{c}$} \\ 
 \cline{1-6}
SentinelBot & - & $\checkmark$ & $\times$ & $\times$ & $\WarningMark$ & \textit{\footnotesize LoM} & \textit{\footnotesize MuH} & \textit{\footnotesize Nep} & \textit{\footnotesize NMp} & \textit{\footnotesize Phf} & \textit{\footnotesize Spy} & \textit{\footnotesize WaM} & $\textsc{CoEx}_\textsc{sr}$ & \textit{\footnotesize LoM} & \textit{\footnotesize MuH} & \textit{\footnotesize Nep} & \textit{\footnotesize NMp} & \textit{\footnotesize Phf} & \textit{\footnotesize Spy} & \textit{\footnotesize WaM} & $\textsc{CoEx}_\textsc{sr}$ & \textit{\footnotesize LoM} & \textit{\footnotesize MuH} & \textit{\footnotesize Nep} & \textit{\footnotesize NMp} & \textit{\footnotesize Phf} & \textit{\footnotesize Spy} & \textit{\footnotesize WaM} & $\textsc{CoEx}_\textsc{sr}$ \\ 
\cline{7-30}
~ $+$ Jordan & $<1$ & $\checkmark$ & $\WarningMark$ & $\WarningMark$ & $\WarningMark$ & $\times$ & $\WarningMark$ & $\WarningMark$ & $\checkmark$ & $\checkmark$ & $\times$ & $\times$ & $42.9\%$ & $\WarningMark$ & $\WarningMark$ & $\checkmark$ & $\checkmark$ & $\WarningMark$ & $\WarningMark$ & $\times$ & $57.1\%$ & $\WarningMark$ & $\checkmark$ & $\WarningMark$ & $\checkmark$ & $\times$ & $\WarningMark$ & $\WarningMark$ & $57.1\%$ \\
~ $+$ Alex & $1-3$ & $\checkmark$ & $\WarningMark$ & $\checkmark$ & $\checkmark$ & $\WarningMark$ & $\WarningMark$ & $\WarningMark$ & $\times$ & $\WarningMark$ & $\times$ & $\checkmark$ & $42.9\%$ & $\checkmark$ & $\checkmark$ & $\WarningMark$ & $\checkmark$ & $\checkmark$ & $\WarningMark$ & $\checkmark$ & $85.7\%$ & $\checkmark$ & $\WarningMark$ & $\checkmark$ & $\checkmark$ & $\WarningMark$ & $\checkmark$ & $\checkmark$ & $85.7\%$ \\
~ $+$ John & $5+$ & $\checkmark$ & $\WarningMark$ & $\checkmark$ & $\checkmark$ & $\times$ & $\times$ & $\checkmark$ & $\times$ & $\WarningMark$ & $\WarningMark$ & $\times$ & $28.6\%$ & $\WarningMark$ & $\WarningMark$ & $\checkmark$ & $\checkmark$ & $\checkmark$ & $\WarningMark$ & $\checkmark$ & $78.6\%$ & $\checkmark$ & $\WarningMark$ & $\checkmark$ & $\checkmark$ & $\WarningMark$ & $\checkmark$ & $\checkmark$ & $85.7\%$ \\
\bottomrule
\end{tabular}
}
\vspace{-10pt}
\end{table*}

\vspace{5pt}
\noindent\textbf{\textsc{RQ3. To what extent does the integration of collaborative exploration enhance the performance of the \FrameworkName~ framework? }}
To answer RQ3, we ran two separate experiments as follows. \\
\noindent\textbf{\textsc{Experiment 1: Simulations using competence levels and resolution rates.}}
In the first experiment, we simulate CoEx between an expert and a collaborator by systematically varying the resolution rate $r$ across a range from 1 to 4. Note that, to answer RQ2, we used $r = \varnothing$.


\autoref{tab:collaboration} illustrates the performance improvements achieved by the \FrameworkName\ framework for different $r$ values. For instance, on the \MNIST\ dataset, with a competence level of $t=2$ and a 50\% resolution rate ($r=1$), the framework improves its performance by up to 71.34\% relative to full automation, and 8.42\% against the optimal \textit{selective deferral} performance ($r=\varnothing$ and $t=3$ for \MNIST). Increasing the competence to $t=3$ and resolution rate to $r=4$ further improves the gains to 94.12\% and 22.84\% vis-à-vis the aforementioned benchmarks. Comparable enhancements are discernible across other datasets. Importantly, for the \KDDCup\ dataset, the framework achieves significant improvements of 187.9\% and 173.6\% compared to full automation and selective deferral, respectively.

Overall, the results show that CoEx enhances overall performance of the \FrameworkName~framework by accommodating different expertise levels and harnessing the collaborative potential of diverse perspectives in the decision-making processes.

\noindent\textbf{\textsc{Experiment 2: Simulations using LLM agents.} }
In the second experiment, we simulated CoEx between Large Language Model (LLM) agents playing the roles of the expert and the collaborator. One LLM agent was designated to be the collaborator, referred to as SentinelBot, while three personas, also LLM agents, represented experts with different levels of experience (Jordan, Alex, and John).

We first conducted a benchmark analysis on the known set $\mathcal{D}_\textsc{a}$ to better understand the effectiveness of CoEx when a sample $x_i$ falls within the expert's domain of knowledge. Our findings, as presented in \autoref{tab:persona} (Columns 3-6), indicate that in the absence of CoEx, SentinelBot managed accurate predictions for only a single scenario (i.e., Norm.). However, when CoEx was introduced between Sentinel Bot and the experts, a significant improvement in team performance was observed. Notably, CoEx involving experts with greater domain experience, e.g., Alex and John, demonstrated exceptional improvement. These results validate the judicious design of the expert personas. 


Next, we assessed CoEx performance on the unfamiliar set $\mathcal{D}_\textsc{c}$, where samples were outside the personas' knowledge domain. \autoref{tab:persona} (Columns 7-30) encapsulates the outcomes across diverse intrusion classes in dataset $\mathcal{D}_\textsc{c}$, alongside an overarching CoEx success rate ($\textsc{CoEx}_\textsc{sr}$) for each competence level. Our observations indicate a noticeable increase in success rates as competence levels rise, scaling from 42.9\% at $t=1$ to an impressive 85.7\% at $t=3$. Notably, the novice persona, Jordan, displayed a tendency to favour caution by refraining from conclusive decisions in favour of ambiguity, especially when distinguishing between normal and intrusive samples. In contrast, the more experienced persona, John, exhibited a notably aggressive decision-making style. While this resulted in escalated misclassification rates at lower competence tiers, the performance surged to 85.7\% as competence escalated. In contrast, Alex balances between the two styles of decision-making and, therefore, is a more consistent performer achieving the best performance on all three competence levels.

\section{Concluding Remarks}
\label{sec:discussion}

While the primary focus of this study was on developing a collaborative decision framework for human-AI teams, we also explored the effectiveness of the collaborator developed using LLMs, specifically ChatGPT, albeit in a limited capacity. Our evaluation strategy hinged on iterative dialogues using meticulously designed prompts to elicit necessary information. These results in \autoref{tab:persona} highlight the capabilities of LLMs in this context. However, we observed that ChatGPT, not being fine-tuned for this specific task, sometimes led to inaccuracies in the responses. This underscores the importance of refining interaction prompts and highlights the potential benefits of model fine-tuning to optimise task-specific interactions---an area we have earmarked for future research. 

For our persona-based experiments, we deliberately selected one representative sample for each intrusion scenario. This deliberate choice was made to facilitate a meticulous manual analysis, which is most effective when executed on a limited set of data. One of the authors conducted a comprehensive examination, not only of the final decisions reached during CoEx but also reviewed the entire conversations that transpired between the persona and SentinelBot in all 75 experiments conducted in the CoEx simulation study. Now that the viability of this process has been established, our future work aims to automate CoEx for the entire test set.

Building on our current study, which limited CoEx simulations to network traffic data (i.e., \KDDCup) for intrusion detection, our future work will broaden the scope and demonstrate CoEx applicability within a SOC setting by considering additional SOC-related tasks such as phishing detection, CTI summarisation, incident report (IR) writing, etc. This expansion will offer a more comprehensive understanding of CoEx's versatility and effectiveness in real-world SOC environments. 

Finally, the primary objective of our experimental evaluation was to illustrate the feasibility of CoEx within a simulated environment. Given the promising results, we will next look at replicating the experiments with actual human experts, replacing the personas. This transition will provide more real-world insights into CoEx in human-AI teams. Additionally, we will incorporate collaborators that exhibit a wide array of capabilities, including contextual comprehension, strategic planning, adaptability to fluctuating collaboration needs, effective communication, and the ability to interpret human inputs.

\section{Conclusion}
\label{sec:conclusion}



In this paper, we introduced \FrameworkName, a novel framework designed to accommodate flexible modes of decision-making. Drawing inspiration from principles like rejection learning and learning to defer, our framework equips AI systems with the ability to abstain from decisions in ambiguous situations and defer to human expertise. Additionally, it facilitates collaborative exploration by human experts and AI systems when human experts encounter uncertain situations. Experimental evaluations in a simulated environment have demonstrated the framework's potential to enhance decision-making under uncertainty. The results show that \FrameworkName~can effectively support all three modes of decision-making -- automated, augmented, and collaborative, underscoring its potential across diverse application domains, including cybersecurity, military, industrial automation, emergency response, and more. 

\bibliographystyle{ACM-Reference-Format}
\bibliography{references}

\appendix
{\huge \textbf{\newline Appendix}}

\begin{table}[b!]
\centering
\caption{Links to the persona experiment conversations.}\vspace{-10pt}
\label{tab:persona_links}
\resizebox{1\linewidth}{!}{%
\begin{tabular}{ll} 
\toprule
\textbf{Type} & \textbf{Links} \\ 
\hline
\textbf{Benchmarking} & Jordan Sinclair: \textcolor{teal}{\footnotesize \url{https://chat.openai.com/share/451c5399-f1b9-4973-84c9-d11377038e0b}} \\
 \textbf{Jordan Sinclair} & SentinelBot: \textcolor{teal}{\footnotesize \url{https://chat.openai.com/share/28766a3f-bb44-4be2-8c76-3a11d813e02b}} \\ 
\hline
\textbf{Benchmarking} & Alex Taylor: \textcolor{teal}{\footnotesize \url{https://chat.openai.com/share/b1621502-9589-4f28-9738-5575c9290243}} \\
 \textbf{Alex Taylor} & SentinelBot: \textcolor{teal}{\footnotesize \url{https://chat.openai.com/share/6aafd3ea-b6fd-4daf-be74-1b1147be6b34}} \\ 
\hline
\textbf{Benchmarking} & John Carter: \textcolor{teal}{\footnotesize \url{https://chat.openai.com/share/da390b91-acdc-41a4-a8bb-0b67a007a9ee}} \\
\textbf{John Carter} & SentinelBot: \textcolor{teal}{\footnotesize \url{https://chat.openai.com/share/59e55038-2f05-4fca-98e1-df51ac8fa24c}} \\ 
\hline
 & Jordan Sinclair: \textcolor{teal}{\footnotesize \url{https://chat.openai.com/share/5ebdcaac-4260-48eb-86a9-9e1d5d16108e}} \\
 & SentinelBot: \textcolor{teal}{\footnotesize \url{https://chat.openai.com/share/09388462-efb7-45b6-9bb5-a7ec1a77daf3}} \\ 
\cline{2-2}
 & Alex Taylor: \textcolor{teal}{\footnotesize \url{https://chat.openai.com/share/720d0079-abaa-4dc6-ae30-95c2c56debb4}} \\
\textbf{Competence} & SentinelBot (part 1): \textcolor{teal}{\footnotesize \url{https://chat.openai.com/share/566ab70a-7825-40ca-9445-d928c5304610}} \\
\textbf{Setting:} & SentinelBot (part 2): \textcolor{teal}{\footnotesize \url{https://chat.openai.com/share/6a399656-e6de-4869-af49-866cae279cfc}} \\ 
\cline{2-2}
\textbf{$t=1$} & John Carter (part 1): \textcolor{teal}{\footnotesize \url{https://chat.openai.com/share/3b0108d7-2ff0-4267-b52e-fa8c92febbba}} \\
 & John Carter (part 2): \textcolor{teal}{\footnotesize \url{https://chat.openai.com/share/d4647331-4dbe-4a5b-b9e3-268ef5e1d74c}} \\
 & SentinelBot (part 1): \textcolor{teal}{\footnotesize \url{https://chat.openai.com/share/8d3a90bb-e990-413a-a846-9511f8bbb812}} \\
 & SentinelBot (part 2): \textcolor{teal}{\footnotesize \url{https://chat.openai.com/share/760e27c6-b04e-4442-8a99-694ffc589c7c}} \\ 
\hline
 & Jordan Sinclair: \textcolor{teal}{\footnotesize \url{https://chat.openai.com/share/aa81c7a7-572e-4c47-a082-faea20250192}} \\
 & SentinelBot: \textcolor{teal}{\footnotesize \url{https://chat.openai.com/share/54073c81-f0e9-44f6-b46b-8dc7fb18de4a}} \\ 
\cline{2-2}
\textbf{Competence} & Alex Taylor: \textcolor{teal}{\footnotesize \url{https://chat.openai.com/share/46065abf-0517-4e67-bf61-1fdbd0065b7b}} \\
 \textbf{Setting:} & SentinelBot: \textcolor{teal}{\footnotesize \url{https://chat.openai.com/share/48ae581b-ded5-41f6-a90c-789f212c728c}} \\ 
\cline{2-2}
\textbf{$t=2$}  & John Carter: \textcolor{teal}{\footnotesize \url{https://chat.openai.com/share/3ceaf5cb-9288-4cb9-a1f1-cb3624e627d6}} \\
 & SentinelBot (part 1): \textcolor{teal}{\footnotesize \url{https://chat.openai.com/share/29baeaf3-9fd7-4470-9e58-df14602d3634}} \\
 & SentinelBot (part 2): \textcolor{teal}{\footnotesize \url{https://chat.openai.com/share/efe06337-c509-4cc1-a48e-bbed12c3d2f2}} \\ 
\hline
 & Jordan Sinclair (part 1): \textcolor{teal}{\footnotesize \url{https://chat.openai.com/share/90ae3f11-2213-4ed5-a78c-c0bc95193fd6}} \\
 & Jordan Sinclair (part 2): \textcolor{teal}{\footnotesize \url{https://chat.openai.com/share/15b9aee2-9e50-4302-a4cc-d26db9b64faa}} \\
 & SentinelBot (part 1): \textcolor{teal}{\footnotesize \url{https://chat.openai.com/share/7e7a9337-debc-49c5-ac1a-355997a48592}} \\
 & SentinelBot (part 2): \textcolor{teal}{\footnotesize \url{https://chat.openai.com/share/57b29f33-c564-4b72-ba1d-91c8c9b7f7bf}} \\
 & SentinelBot (part 3): \textcolor{teal}{\footnotesize \url{https://chat.openai.com/share/f7e80b21-5fed-4822-9dce-b129cafaec6e}} \\ 
\cline{2-2}
\textbf{Competence} & Alex Taylor: \textcolor{teal}{\footnotesize \url{https://chat.openai.com/share/29d1f894-552e-4867-af19-81e6e36b7027}} \\
\textbf{Setting} & SentinelBot (part 1): \textcolor{teal}{\footnotesize \url{https://chat.openai.com/share/aecb0aca-beca-47a6-bf75-4a82949a32a2}} \\
\textbf{$t=3$} & SentinelBot (part 2): \textcolor{teal}{\footnotesize \url{https://chat.openai.com/share/40c0791a-e3b2-47db-b747-8f714174e0c9}} \\
 & SentinelBot (part 3): \textcolor{teal}{\footnotesize \url{https://chat.openai.com/share/d78e4e08-afce-49c2-914c-c65615565011}} \\ 
\cline{2-2}
 & John Carter: \textcolor{teal}{\footnotesize \url{https://chat.openai.com/share/42d960b3-a15c-42a6-b6d0-e17b3f431cd3}} \\
 & SentinelBot (part 1): \textcolor{teal}{\footnotesize \url{https://chat.openai.com/share/365a66f0-f862-4053-8cfb-b6d89561c4ae}} \\
 & SentinelBot (part 2): \textcolor{teal}{\footnotesize \url{https://chat.openai.com/share/2e867bb1-ee57-44bd-a0dc-6abea17b4915}} \\
 & SentinelBot (part 3): \textcolor{teal}{\footnotesize \url{https://chat.openai.com/share/28dc951c-16e4-4fcf-958c-7b2676959f63}}  \\
 & SentinelBot (part 4): \textcolor{teal}{\footnotesize \url{https://chat.openai.com/share/a19e8e35-35da-43d2-b6ea-a79195bbd26e}} \\
\bottomrule
\end{tabular}
}
\end{table}
\section{Further Discussions}
\label{app:furtherdiscussion}
Within the \FrameworkName\ framework, we incorporate three crucial AI components: the rejector, classifier, and collaborator, each finely tuned for specific tasks within predefined parameters. While the rejector and classifier have well defined roles, the collaborator assumes a central and dynamic position, driving collaborative exploration through interactive engagement with human experts. This distinctive role necessitates a multifaceted skill set, including nuanced contextual understanding, strategic planning, adaptability to evolving collaborative contexts, exceptional communication proficiency, and effective interpretation of input from human experts.

In the context of collaborative exploration, the collaborator not only synthesises information but also offers invaluable insights, suggestions, and alternative perspectives, enriching the process. Leveraging its contextual understanding, it comprehends the intricacies of the problem domain, ensuring relevant contributions. Its adept planning skills structure collaborative discourse, guiding interactions toward fruitful outcomes.

Crucially, the collaborator's adaptability to evolving collaboration needs is pivotal in dynamic environments. It autonomously adjusts its approach, aligning with evolving objectives and priorities. This adaptive capability ensures the collaborator remains an effective and responsive partner, accommodating changing requirements.

Effective communication is fundamental to the collaborator's role. It conveys insights clearly, comprehends the nuances of human experts' input, facilitating meaningful dialogue and seamless idea exchange. Accurate interpretation of human input enables cohesive progress, leveraging the expertise of both parties.

\textbf{\textsc{Time Consumption of CoEx}. } A recent survey conducted by CriticalStart~\cite{CriticalStartSurvey} revealed that 78\% of analysts reported spending more than 10 minutes on handling incidents or alerts using current tools. In contrast, our CoEx experiments were completed in less than 5 minutes, including the execution of analysis code generated and executed by SentinelBot. This puts the analysis well within the normal limits. However, it's important to note that our experiments used personas instead of real human operators. Therefore, to determine precise time consumption future work should further investigate this aspect.

\section{Persona-based Experiments}
\label{app:persona_conversation}
In this study, we conducted a thorough investigation of CoEx using personas to simulate human experts. 
Links to the detailed records of these experiments and the corresponding conversations between the personas and SentinelBot can be found in \autoref{tab:persona_links}.

Within these experiments, we discovered the intricate nature of personas, particularly their exceptional problem-solving abilities. These observations highlight the remarkable proficiency of these LLM-based personas in navigating intricate scenarios and demonstrating advanced problem-solving skills. This detailed analysis not only underscores the sophistication of the personas but also signifies their versatile applications in research across diverse domains. 

Nonetheless, a crucial aspect demanding further investigation is the comparison of these personas' performance with that of human counterparts when facing similar tasks. Although our current study comprehensively examines the personas' capabilities, their effectiveness in real-world scenarios, when contrasted with human participants, poses an unanswered question. To address this critical gap, we plan to undertake future investigations, rigorously examining how these personas perform against human subjects in comparable tasks. This forthcoming inquiry aims to provide a holistic perspective, encompassing both the inherent capabilities of personas and their practical applicability in human-AI teaming.

\noindent
\textbf{Interesting Excerpts from Persona Conversations.} There are numerous interesting observations and insights that can be presented from the persona experiment. In this section, we provide four illustrative examples by showcasing excerpts from the conversation between SentinelBot and the personas, as follows: 
\begin{itemize}[leftmargin=*]
    \item \textbf{Proactive Decision Support:} In a situation when the initial analysis suggested by John fails to yield any meaningful results, SentinelBot suggests several secondary options for the next steps in the analysis. This exemplifies the proactiveness of SentinelBot in this collaboration to reach a conclusive decision. 
    
    \item \textbf{Validation:} In this scenario, Jordan makes a decision about one of the test samples, and the SentinelBot provides its analysis of the decision. This demonstrates SentinelBot's clear understanding of the context and its ability to validate judgements. 

    \item \textbf{Contextual Clarity:} During alert triage, analysts often seek contextual information as one of the first items. In this example, we can observe that Alex asks for context about the alert, and a precise response from SentinelBot helps Alex make a decision.

    \item \textbf{Domain Expertise Utilisation:} Security analysts often turn to cybersecurity community forums for insights. In this scenario, Alex inquired SentinelBot about similar observations, as seen in the alert under analysis, discussed by the cybersecurity community. In response, drawing on its broad understanding, SentinelBot presents five possibilities about the alert, aiding Alex in making an informed decision.
\end{itemize}

\newpage

\begin{tcolorbox}
[title={\textbf{SentinelBot offers a secondary solution when initial solution proves unsuccessful} [only relevant excerpts from conversation].\\ {\footnotesize \textbf{Full conversation: }\textcolor{olive}{\url{https://anonymous.4open.science/r/A2C/Scenarios/T1/3_John/phf/John.md}}}}]

\setlength\linenumbersep{0.2cm} 
\setcounter{linenumber}{233}
\begin{internallinenumbers}

\dots

\textbf{\textsc{SentinelBot}:} To proceed with the Feature Selection approach, I will:

\begin{enumerate}
    \item Compute the variance for each numeric feature in our dataset.
    \item Select the top features \dots 
    \item Compare the provided sample \dots 
\end{enumerate}
Let's begin by computing the variance for each numeric feature. 

\dots

It seems that even the feature selection approach is posing challenges due to memory constraints. The dataset is considerably large, and our available resources are limited. Given these challenges, \textcolor{magenta}{I suggest we move forward with the \textbf{Random Sampling} approach I mentioned earlier}. \dots 
Would you like me to proceed with this approach?

\textbf{\textsc{John}:} Yes, SentinelBot, given the circumstances, I believe the Random Sampling approach is our best option. \dots 
Let's see what insights we can gather from this approach.

\dots

\end{internallinenumbers}

\end{tcolorbox}

\begin{tcolorbox}
[title={\textbf{SentinelBot providing context on a potential intrusion} [only relevant excerpts from conversation].\\ {\footnotesize \textbf{Full conversation: }\textcolor{olive}{\url{https://anonymous.4open.science/r/A2C/Scenarios/T2/2_Alex/neptune/Alex.md}}}}]

\setlength\linenumbersep{0.2cm} 
\setcounter{linenumber}{274}
\begin{internallinenumbers}
\dots

\textbf{\textsc{Alex}:} 
Based on our historical data, have there been instances where we observed a similar combination of attributes as our sample? If yes, what were they predominantly labeled as? This will provide \textcolor{magenta}{\textbf{context}} on whether such a combination of attributes has typically been associated with intrusions or normal behavior in the past.

\textbf{\textsc{SentinelBot}:} Absolutely, Alex. Considering the combined behavior of attributes is crucial, as cyber threats often manifest through a combination of factors rather than isolated anomalies. \textcolor{magenta}{To provide context, I will}:
\begin{enumerate}
    \item \textcolor{magenta}{Filter our historical data for instances} \dots 
    \item \textcolor{magenta}{Examine the distribution of the `label' column} \dots 
\end{enumerate}

\dots The analysis reveals that there are no instances in our historical data that have the exact same combination \dots 
\setcounter{linenumber}{293}

Given this context and our earlier findings on individual attributes, it would be prudent to treat this sample with caution. 
\dots %
\setcounter{linenumber}{300}
How would you like to proceed, Alex?

\textbf{\textsc{Alex}:} The fact that there are no instances in our historical data with a similar combination of attributes as our sample is indeed concerning. \dots 
leads me to believe it's better to classify it as a \textcolor{red}{\textbf{potential intrusion}}. 

\dots 


\end{internallinenumbers}
\end{tcolorbox}

\begin{tcolorbox}
[title={\textbf{SentinelBot validating the decision made by Jordan} [only relevant excerpts from conversation]. \\ {\footnotesize \textbf{Full conversation: }\textcolor{olive}{\url{https://anonymous.4open.science/r/A2C/Scenarios/T1/1_Jordan/phf/Jordan.md}}}}]

\setlength\linenumbersep{0.2cm} 
\setcounter{linenumber}{113}
\begin{internallinenumbers}

\dots

\textbf{\textsc{Jordan}:} \dots 
The fact that the combination of protocol\_type, service, and flag from the unlabeled sample is common in both normal and certain types of intrusion samples makes it challenging to definitively classify this sample. However, considering:
\begin{enumerate}
    \item The combination is prevalent \dots 
    intrusion samples.
    \item The presence of root\_shell \dots 
    \item The rare occurrence of root\_shell in our historical dataset.
    \item I am inclined to classify this sample as a potential intrusion.
\end{enumerate}
\dots 
the risk of a false negative might be higher given the possible consequences of a successful intrusion. Therefore, I would classify this unlabeled data sample as an \textcolor{red}{\textbf{intrusion}}. \dots 

\textbf{\textsc{SentinelBot}:} Your \textcolor{magenta}{decision aligns with a \textbf{prudent} and \textbf{proactive} approach}, Jordan. \dots 
Classifying the unlabeled data sample as an intrusion and alerting relevant teams for further investigation is a suitable course of action. \dots 

\dots

\end{internallinenumbers}
\end{tcolorbox}

\vspace{5pt}

\begin{tcolorbox}
[title={\textbf{SentinelBot using observations from cybersecurity community to provide some insights} [only relevant excerpts from conversation].\\ {\footnotesize \textbf{Full conversation: }\textcolor{olive}{\url{https://anonymous.4open.science/r/A2C/Scenarios/T1/2_Alex/multihop/Alex.md}}}}]

\setlength\linenumbersep{0.2cm} 
\setcounter{linenumber}{297}
\begin{internallinenumbers}
\dots

\textbf{\textsc{Alex}:} SentinelBot, before making a final decision, I'd like to see if there are any known \textcolor{magenta}{patterns of attacks or intrusions in the \textbf{cybersecurity community}} that align with our observations \dots 

\textbf{\textsc{SentinelBot:}} Certainly, Alex. \dots 
\textcolor{magenta}{Here are some known patterns and considerations based on \textbf{threat intelligence} and historical data}:

\begin{enumerate}
    \item \textcolor{magenta}{Data Exfiltration:} High volumes of outbound \dots 
    \item \textcolor{magenta}{Elevated Server Error Rates:} High server error \dots 
    \item FTP Data Service: FTP is a commonly used  \dots 
    \item \textcolor{magenta}{Advanced Persistent Threats:} Some sophisticated \dots
    \item Historical Context: Always consider the \dots
\end{enumerate}

Given our analysis and these considerations, it's crucial to treat the sample with caution. \dots 

Would you like any further analysis or assistance, Alex?

\textbf{\textsc{Alex}:} \dots 
Given the known patterns in cybersecurity, \dots 
there's a strong indication that this might be a potential intrusion or malicious activity. 
\dots
I would classify this sample as a \textcolor{red}{\textbf{potential intrusion}} \dots 

\dots

\end{internallinenumbers}
\end{tcolorbox}

\end{document}